\documentclass[twocolumn,showpacs,preprintnumbers,amsmath,amssymb,floatfix]{revtex4}
\usepackage[dvips]{graphicx}% Include figure files

\usepackage{dcolumn}% Align table columns on decimal point
\usepackage{bm}% bold math
\usepackage{color}

\begin{document}

\title{Polarization ququarts}
\author{Yu. I. Bogdanov}
\affiliation{Institute of Physics and Technology, Russian Academy of Science}
\author{E. V. Moreva}
\affiliation{Moscow Engineering Physics Institute (State
University)Russia}
\author{G. A. Maslennikov}
\affiliation{National University of Singapore, Singapore}
\author{R. F. Galeev}
\author{S. S. Straupe, and S. P. Kulik}
\affiliation{Faculty of Physics, Moscow State Institute, Russia}

\date{\today}
\begin{abstract}

We discuss the concept of polarization states of four-dimensional
quantum  systems based on frequency non-degenerate biphoton field.
Several quantum tomography protocols were developed and implemented
for measurement of an arbitrary state of ququart. A simple method
that does not rely on interferometric technique is used to generate
and measure the sequence of states that can be used for quantum
communication purposes.
\end{abstract}
\pacs{42.50.-p, 42.50.Dv, 03.67.-a} \maketitle

\section{Introduction}

  Recently multi-dimensional quantum systems (quantum dits, or qudits)
attract much attention in context of quantum information and
communication. It is partly caused by fundamental aspects of quantum
theory since usage of qudits allows one to violate Bell-type
inequalities strongly than with two dimensional systems (qubits)
\cite{quBell}. Much interest in qudits also comes from the
application point, especially from applied quantum key distribution.
Multilevel systems are proved to be more robust against noise in the
transmission channel, although measurement and preparation
procedures of such systems seem to be much more technically
complicated than in the case of qubits. Different aspects of the
security of qudits-based protocols have been analysed
\cite{QKD_Security} including those related to the reduced set of
bases \cite{Tittel}. Lately a proof-of-principal realization of a
QKD protocol with qudits \cite{Walborn:05} and with entangled
qutrits ($D=3$, where $D$ is dimensionality of the Hilbert space)
\cite{QKD_Zeilinger:05} have been demonstrated. For the last several
years elegant experiments were performed where different kinds of
optical qudits were introduced. Most of them are based on states of
light emitted via spontaneous parametric down-conversion (SPDC).
There are qudits obtained using spatial degrees of freedom: with
either single photons \cite{Walborn:05} or twin photons
\cite{Howell:05}; orbital angular momentum of photons \cite {OAM};
time-bin tecnique \cite{timebin}; multi-armed interferometers
\cite{Thew:04}; postselection of polarization qutrits from
four-photon states \cite{antia:01}, and polarization states of
single-mode biphotons \cite{ourPRL:04}.

It seems that among manifold manners of qudits preparation only
the method based on polarization states of single-mode biphotons
admits preparation of any qutrit state (pure or mixed) on demand
within one set-up and guarantees complete control over the state
with high accuracy \cite{ourPRL:04,ourPRA:04}. However, this
method does not allow to create entangled qutrits. Another
limitation of polarization qutrits relates to the fundamental lock
for their basic states transformations using SU(2) optical
elements, like retardation plates, rotators etc. Meanwhile it is
these transformations that would be extremely useful for quantum
communication purposes.

In this paper we present the results of experimental preparation
and measurement carried out with polarization based ququarts or
quantum systems with dimensionality $D=4$. The paper is organized
as follows. In Sec.II we discuss the main properties of ququarts
based on the polarization state of two-photon (biphoton) field.
Such operational notions as coherence matrix, polarization degree
are introduced. The criterium of separability for qubits forming
ququarts is discussed as well. Sec.III is devoted to different
experimental implementations of biphoton-ququarts preparation and
their measurement. In Sec.IV we consider a specific set of ququarts
which can be used in practice for quantum key distribution.

\section{Polarization ququarts and their properties}
\subsection{Polarization states of biphoton field}
Biphoton field generated via spontaneous parametric
down-conversion process is represented by a coherent mixture of
two-photon Fock states and the vacuum state~\cite{klbook}:

\begin{equation} \Psi = | {vac} \rangle + \frac{1}{2}\sum\limits_{\vec{k_i}\vec{k_s}} {F_{\vec{k_i},\vec{k_s} }
| {1_{\vec{k_i} } ,1_{\vec{k_s} } } \rangle } ,\end{equation} where
$| {1_{\vec{k_i} } ,1_{\vec{k_s} } } \rangle $ denotes the state
with one (idler) photon in the mode $\vec{k_{i}}$ and one (signal)
photon in the mode $\vec{k_{s}}$. Since squared modulus of the
coefficient $F_{\vec{k_i} ,\vec{k_s} } $ gives probability to
register two photons in modes $\vec{k_{i}}$ and $\vec{k_{s}}$ it is
called the biphoton amplitude ~\cite{kllaser}.
The pure polarization
state of down-converted light field, which is often refered as
two-photon field (or biphotons), can be written as follows:
\begin{equation}
|\Psi\rangle=c_{1}|2_{H},0_{V}\rangle+c_{2}|1_{H},1_{V}\rangle+c_{3}|1_{V},1_{H}\rangle+c_{4}|0_{H},2_{V}\rangle.
\label{eq:stategen}
\end{equation}
Here $c_i=|c_i|e^{i\phi_i}$, $\sum\limits_{i = 1}^4 {| {c_i} |^2 =
1} $ are complex probability amplitudes. The first ($n$) and second
($m$) place in kets corresponds to number of distinct photons with
definite (either horizontal $H$ or vertical $V$) polarization, with
total photon number $m+n=2$. If down converted photons have equal
frequency and momentum, then a ququart state (\ref{eq:stategen})
converts to a qutrit state, i.e. middle terms in (\ref{eq:stategen})
become indistinguishable: $\vec {k}_s \approx \vec {k}_i $, $\omega
_s \approx \omega _i $ and $\omega _s + \omega _i = \omega _p $,
where $\omega _p $ is the laser pump frequency. In this case any
arbitrary pure polarization state of biphoton field can be expressed
in terms of three complex amplitudes $c_1^{\prime} ,c_2^{\prime} ,$
and $c_3^{\prime} $:

\begin{equation} | c^{\prime}\rangle = c_1^{\prime} | {2,0} \rangle + c_2^{\prime} | {1,1} \rangle + c_3^{\prime} |
{0,2} \rangle , \end{equation}with $c_i^{\prime} = | {c_i^{\prime} }
|\exp \{ {i\varphi _i } \}$, $\sum\limits_{i = 1}^3 {| {c_i^{\prime}
} |^2 = 1} $. The vector $| c^{\prime} \rangle = ( {c_1^{\prime}
,c_2^{\prime} ,c_3^{\prime} } )$ represents a three-state system or
a qutrit. Figure 1 presents the photograph picture of two
dimensional SPDC spectrum taken in coordinates wavelength-angle. By
convention the left-and-upper side in respect with doubled pump
wavelength belongs to "signal" range whereas the right-and-down side
belongs to "idler" one. The dashed lines limit the central part
of the spectrum which corresponds to the frequency degenerate and
collinear regime when photons forming the biphoton have
approximately the same wavelengths and propagate along the pump. To
select this part pinholes and/or narrow-band filters are typically
used. It is that part of biphoton spectrum, that contributes to
qutrits.
\begin{figure}[!ht]
\includegraphics[width=0.4\textwidth]{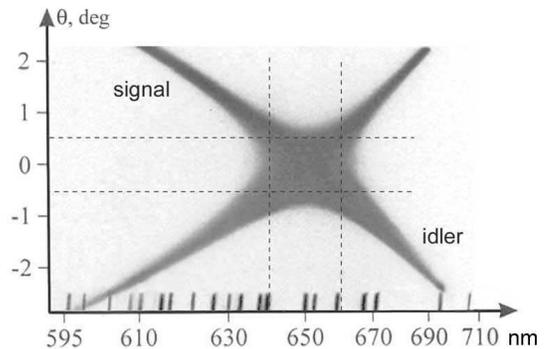}
\caption{Photograph of lithium iodate ($LiIO_{3}$)
frequency-angular spectrum taken with helium cadmium laser
operated at 325nm. Angle between crystal optical axis and laser
pump is $59.22^\circ$. Dashed lines select the part of the
spectrum (frequency degenerate, collinear regime of SPDC)
contributed to biphotons-qutrits. Vertical lines at the bottom
belong to the spectrum of neon which serves as a frequency
reference.}
\end{figure}
In order to distinguish between middle terms in (\ref{eq:stategen})
one must induce the distinguishability between down converted
photons either in frequency, momentum or detection time. In
experiments described in this paper we chose collinear
non-degenerate regime of SPDC so twin photons that form a biphoton
were having different frequencies and propagate along the same
direction. The appropriate two-dimensional spectrum is shown in
Figure 2. To take this spectrum it is enough to tilt the crystal by
a small angle with respect to the orientation that is used for
degenerate regime. Two dashed rectangles indicate the
angular-frequency ranges of spectrum contributing to ququarts.
\begin{figure}[!ht]
\includegraphics[width=0.4\textwidth]{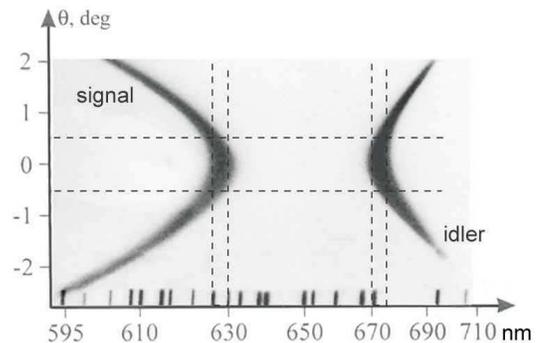}
\caption{Non-degenerate regime of SPDC. Angle between crystal
optical axis and laser pump is $58.97^\circ$. Dashed lines select
the part of the spectrum contributed to ququarts.}
\end{figure}
The state (\ref{eq:stategen}) can be thus rewritten in the form:
\begin{equation}
|\Psi\rangle=c_{1}|H_{1},H_{2}\rangle+c_{2}|H_{1},V_{2}\rangle+
c_{3}|V_{1},H_{2}\rangle+c_{4}|V_{1},V_{2}\rangle.
\label{eq:state}
\end{equation}

Here $|H_{{1(2)}}\rangle\equiv a_{1(2)}^{\dagger}|vac\rangle$,
$|V_{{1(2)}}\rangle\equiv b_{1(2)}^{\dagger}|vac\rangle$, where
$a_{1(2)}^{\dagger}$, $b_{1(2)}^{\dagger}$ are the creation
operators of down-converted photons with central wavelength
$\lambda_{1}$($\lambda_{2}$) in horizontal and vertical polarization
mode. The alternative way to represent a ququart might be by
introducing distinguishability between spatial modes while keeping
the same central wavelengths of photons: $\vec {k}_s \neq\vec {k}_i
$, $\omega _s \approx \omega _i$. The detailed description of
polarization properties of such a state of two qubits has been done
by James and co-authors \cite{Kwiat:01}. The case of ququart with
two frequency modes attracts much attention because for some class
of the tasks it is convenient to operate with states in single
spatial mode.
\subsection{Is ququart a separable or entangled state of two qubits?}
\par Generally, the
state (\ref{eq:state}) can not be written as a direct product of
polarization states of two photons:
\begin{equation}
|\Psi_1\rangle\otimes|\Psi_2\rangle=(p_{1}|H_{1}\rangle+q_1|V_{1}\rangle)\otimes(p_{2}|H_{2}\rangle+q_2|V_{2}\rangle)
\label{eq:facstate}
\end{equation}
That is why it is useful to write down the condition when
(\ref{eq:state}) becomes a separable state. The reduced density
matrix of the subsystem (second photon) can be found by tracing
the joint state over all states of the first photon:
\begin{equation}
\rho_2= Sp_{1}(|\Psi\rangle\langle\Psi|)=\left(
{{\begin{array}{*{20}c}
{|c_{1}|^2+|c_{3}|^2 }\hfill & {{c_{1}c_{2}^*}+ {c_{3}c_{4}^*}}\hfill\\
 {{c_{2}c_{1}^*}+ {c_{4}c_{3}^*}} \hfill & {|c_{2}|^2+|c_{4}|^2 }\hfill \\
 \end{array} }} \right).
\label{eq:rho2}
\end{equation}
It is straightforward to show that the eigenvalues of
(\ref{eq:rho2}) take the meanings $\lambda_{1,2}=0, 1$ and entropy
of each subsystem drops to zero iff
\begin{equation}
c_1c_1^*c_4c_4^*+c_3c_3^*c_2c_2^*-c_2c_1^*c_3c_4^*-c_4c_3^*c_1c_2^*= \\
|c_1c_4-c_2c_3|^2=0.
\label{eq:criterium}
\end{equation}
Thus if $c_1c_4=c_2c_3$ holds, the state (\ref{eq:state}) reduces to
factorized one with definite states of each qubit $|\Psi_1\rangle$,
$|\Psi_2\rangle$ and vice versa. In other words (\ref{eq:criterium})
plays the role of criterium for biphoton-based ququart to be a
separable state of a couple of polarization qubits belonging to two
different modes. Sub-indexes in (\ref{eq:state}) designate either
frequency or spatial modes.
\par As examples of different ququarts we refer to the product states that might be generated from a single non-linear
crystal via SPDC:
\begin{equation}{{\begin{array}{*{20}c}
type I:\hfill & |H_{1}H_{2}\rangle=\left( {{\begin{array}{*{20}c}
{1} \hfill \\
{0} \hfill \\
{0} \hfill \\
{0} \hfill \\
\end{array} }} \right), \hfill & |V_{1}V_{2}\rangle=\left( {{\begin{array}{*{20}c}
{0} \hfill \\
{0} \hfill \\
{0} \hfill \\
{1} \hfill \\
\end{array} }} \right),\hfill\\
type II:\hfill & |V_{1}H_{2}\rangle=\left( {{\begin{array}{*{20}c}
{0} \hfill \\
{0} \hfill \\
{1} \hfill \\
{0} \hfill \\
\end{array} }} \right), \hfill & |H_{1}V_{2}\rangle=\left( {{\begin{array}{*{20}c}
{0} \hfill \\
{1} \hfill \\
{0} \hfill \\
{0} \hfill \\
\end{array} }} \right).\hfill\\
\end{array} }}
\label{eq:basis}
\end{equation}

and maximally entangled Bell-states:
\begin{equation}
|\Psi^{\pm}\rangle=\left( {{\begin{array}{*{20}c}
{0} \hfill \\
{\frac{1}{\surd\bar{2}}} \hfill \\
{\pm\frac{1}{\surd\bar{2}}} \hfill \\
{0} \hfill \\
\end{array} }} \right),\quad
|\Phi^{\pm}\rangle=\left( {{\begin{array}{*{20}c}
{\frac{1}{\surd\bar{2}}} \hfill \\
{0} \hfill \\
{0} \hfill \\
{\pm\frac{1}{\surd\bar{2}}} \hfill \\
\end{array} }} \right),
\label{eq:Bell}
\end{equation}

\par At the end of this section we would like to mention that
biphoton-ququart is not supposed to be mapped on Poincar\'{e}
sphere as a pair of points like it occurs for biphoton-qutrits
\cite{ortogonality, burjetp}. Indeed it is not because the state
(\ref{eq:state}) can not be factorized in the general case.
However, if (\ref{eq:criterium}) holds the sub-set of separable
qubit states admits its visual representation on Poincar\'{e}
sphere.
\subsection{Second order of the field: Stokes parameters}
Polarization properties of a pure state (\ref{eq:state}) can be
described by Stokes parameters which are the mean values of Stokes
operators averaged over the state (\ref{eq:state}). Although the
description of light polarization can be introduced only for the
quasimonochromatic plane waves, it is possible to use $P$-quasispin
formalizm \cite{karasev} to describe the polarization of arbitrary
quantum beams with $n$ modes, frequency or spatial. For
two-frequency and single-spatial mode field, the Stokes parameters
will contain time and frequency dependent terms
$exp(i(\omega_{j}-\omega_{k})t)$ that describe "beats" of frequency
modes and have no connection with light polarization. However, these
terms vanish if one considers the finite detection time that allows
to classically average these "beatings". So in the case of two
frequencies, the Stokes operators are given by the sum of
corresponding operators in two modes

\begin{equation}
\begin{array}{cc}
\langle S_{0}\rangle=\langle
a_{1}^{\dagger}a_{1}+a_{2}^{\dagger}a_{2}+b_{1}^{\dagger}b_{1}+b_{2}^{\dagger}b_{2}\rangle\equiv&\\\langle
S_{0}^{(1)}\rangle+\langle S_{0}^{(2)}\rangle=2; &\\\langle
S_{1}\rangle=\langle
a_{1}^{\dagger}a_{1}+a_{2}^{\dagger}a_{2}-b_{1}^{\dagger}b_{1}-b_{2}^{\dagger}b_{2}\rangle\equiv
&\\\langle S_{1}^{(1)}\rangle+\langle S_{1}^{(2)}\rangle= 2(|c_{1}|^2-|c_{4}|^2); &\\
\langle S_{2}\rangle=\langle
a_{1}^{\dagger}b_{1}+a_{2}^{\dagger}b_{2}+b_{1}^{\dagger}a_{1}+b_{2}^{\dagger}a_{2}\rangle
\equiv&\\\langle S_{2}^{(1)}\rangle+\langle S_{2}^{(2)}\rangle= \\
2Re(c_{1}^{\star}(c_{2}+c_{3})+c_{4}(c_{2}^{\star}+c_{3}^{\star}));
&\\\langle S_{3}\rangle=\langle
a_{1}^{\dagger}b_{1}+a_{2}^{\dagger}b_{2}-b_{1}^{\dagger}a_{1}-b_{2}^{\dagger}a_{2}\rangle
\equiv&\\\langle S_{3}^{(1)}\rangle+\langle
S_{3}^{(2)}\rangle=&\\2Im(c_{1}^{\star}(c_{2}+c_{3})+c_{4}(c_{2}^{\star}+c_{3}^{\star})).
\end{array}
\label{eq:Stokes}
\end{equation}

The same definition applies to the second wavelength $\lambda_{2}$
and we take into account that these operators do not commute for
different frequency modes.

\subsection{Fourth order of the field: Coherence matrix}
Polarization properties of biphoton-ququarts are determined
completely by the analogue of the coherence matrix.  It is a matrix
consisting of ten fourth-order moments of the electromagnetic
field. An ordered set of such moments can be obtained using the
direct product of $2\times{2}$-coherence matrixes for both
polarization qubits forming biphoton:

\begin{equation}
K_4 \equiv(K_2)_1\otimes(K_2)_2=\left( {{\begin{array}{*{20}c}
 A \hfill & E \hfill & F \hfill & G \hfill\\
 {E^ * } \hfill & B \hfill & I \hfill & K \hfill\\
 {F^ * } \hfill & {I^ * } \hfill & C \hfill & L \hfill\\
 {G^ * } \hfill & {K^ * } \hfill & {L^ *} \hfill & D \hfill\\
\end{array} }} \right),
\label{eq:K4}
\end{equation}
where $(K_2)_{1,2}$ are $2\times{2}$-coherence matrixes for single
photons with different frequencies marked with indexes $j = 1, 2$:
\begin{equation}
(K_ 2)_j\equiv\left( {{\begin{array}{*{20}c} \langle
a_{j}^{\dagger}a_{j}\rangle \hfill & \langle
a_{j}^{\dagger}b_{j}\rangle\hfill\\ \langle
a_{j}b_{j}^{\dagger}\rangle \hfill &\langle
b_{j}^{\dagger}b_{j}\rangle\hfill\\
\end{array} }} \right),
\label{eq:K2}
\end{equation}

  The diagonal elements of (\ref{eq:K4}) are formed by real moments, which
characterize the intensity correlation in two polarization modes
$H$ and $V$:

\begin{equation}
{{\begin{array}{*{20}c} A \equiv \langle
a_{1}^{\dagger}a_{2}^{\dagger}a_{1}a_{2}\rangle=|{c_1}|^2, \quad
 B \equiv \langle
a_{1}^{\dagger}b_{2}^{\dagger}a_{1}b_{2}\rangle=|{c_2}|^2,\hfill &\\
 C \equiv \langle
b_{1}^{\dagger}a_{2}^{\dagger}b_{1}a_{2}\rangle=|{c_3}|^2, \quad D
\equiv \langle
b_{1}^{\dagger}b_{2}^{\dagger}b_{1}b_{2}\rangle=|{c_4}|^2.\hfill &\\
\end{array} }}
\label{eq:diag}
\end{equation}

Nondiagonal moments are complex:

\begin{equation}
{{\begin{array}{*{20}c} E \equiv \langle
a_{1}^{\dagger}a_{2}^{\dagger}a_{1}b_{2}\rangle=c_1^*c_2, \quad
 F \equiv \langle
a_{1}^{\dagger}a_{2}^{\dagger}b_{1}a_{2}\rangle=c_1^*c_3,\hfill &\\
 G \equiv \langle a_{1}^{\dagger}a_{2}^{\dagger}b_{1}b_{2}\rangle=c_1^*c_4,
\quad I \equiv \langle a_{1}^{\dagger}b_{2}^{\dagger}b_{1}a_{2}\rangle=c_2^*c_3,\hfill &\\
K \equiv \langle
a_{1}^{\dagger}b_{2}^{\dagger}b_{1}b_{2}\rangle=c_2^*c_4,
\quad L \equiv \langle b_{1}^{\dagger}a_{2}^{\dagger}b_{1}b_{2}\rangle=c_3^*c_4.\hfill &\\
\end{array} }}
\label{eq:nondiag}
\end{equation}

The averaging in (\ref{eq:diag}, \ref{eq:nondiag}) is taken over
the state (\ref{eq:state}). The polarization (reduced) density
matrix of ququart coincides with coherency matrix $K_4$.

\subsection{Polarization degree and transformations of ququarts}
The polarization degree is given by

\begin{equation}
P={\frac{\sqrt{\sum\limits_{k=1}^3{\langle
S_{k}\rangle}^2}}{\langle S_{0}\rangle}}={
\frac{\sqrt{\sum\limits_{k = 1}^3{\langle
S_{k}^{(1)}+S_{k}^{(2)}\rangle}^2}}{\langle
S_{0}^{(1)}+S_{0}^{(2)}\rangle}}.
\label{eq:poldeg}
\end{equation}

This definition of polarization degree is just generalization of
the commonly used classical one. It differs from the definition
suggested in \cite{gunnar}, where it serves as a witness of the
state purity. In the case of polarization-based qutrit states
\cite{ourPRL:04, ourPRA:04}, the polarization degree is
\begin{equation}
P_{3} = \sqrt {|c_1^{\prime}|^2 - |c_3^{\prime}|^2 + 2|
{c_1^{\prime\ast} c_2^{\prime} + c_2^{\prime\ast} c_3^{\prime}}
|^2} \label{eq:poldeg3}
\end{equation}

 with $ c_1^{\prime}=c_1, \sqrt{2}c_2^{\prime}= c_2=c_3,
c_3^{\prime}=c_4$. This quantity was invariant under unitary
polarization transformations, which is reasonably straightforward
\cite{dnk:97}. Indeed it was impossible to prepare all demanded
pure states that can be used, for example, in QKD protocols,
unless one uses interferometric schemes with several nonlinear
crystals \cite{ourPRL:04}, or introduces losses. In particular,
there is no way to transform the basic qutrit state
$|\Psi_{4}^{\prime}\rangle=|V,V\rangle$ with $P$ = 1 into
orthogonal state $|\Psi_{2}^{\prime}\rangle=|H,V\rangle$ with $P$
= 0 using retardation plates \cite{polardegree}. However, in the
case of polarization ququarts, this quantity is no longer
invariant. Indeed the polarization degree (\ref{eq:poldeg}) can be
rewritten as
\begin{equation}
P_{4}={\frac{\sqrt{\sum\limits_{j=1}^2(Tr^2(K_2)_j-2det(K_2)_j)+2\sum\limits_{k
= 1}^3\langle
S_{k}^{(1)}S_{k}^{(2)}\rangle}}{\sum\limits_{j=1}^2Tr(K_2)_j}}.
\label{eq:poldeg4}
\end{equation}
In (\ref{eq:poldeg4}) we used trivial links between coherence
matrices and Stokes parameters \cite{MandelWolf}.

The expression in the round brackets is invariant under unitary
transformations as well as the denominator. The second term under
the square root in numerator can be expanded in terms of
fourth-order moments:
\begin{equation}
\sum\limits_{k = 1}^3\langle
S_{k}^{(1)}S_{k}^{(2)}\rangle=\{TrK_{4}-2(ReI+B+C)\}.
\end{equation}
Since unitary transformations keep number of photons, then
$TrK_{4}$ is invariant. At the same time it is easy to prove that
combination of moments $ReI+B+C$ is not an invariant, so the whole
expression (\ref{eq:poldeg4}) changes under SU(2) transformations.
 As a consequence, the polarization degree changes by applying
local unitary transformations in each frequency mode. This can be
achieved by using dichroic polarization transformers, which act
separately on the photons with different frequencies.
For example, to transform the state
\begin{equation}
|\Psi_{4}\rangle=|V_{\lambda_{1}},V_{\lambda_{2}}\rangle\Rightarrow
|\Psi_{2}\rangle=|H_{\lambda_{1}},V_{\lambda_{2}}\rangle,
\label{eq:vv}
\end{equation}
one needs to use the retardation plate which serves as a half wave
plate at $\lambda_{1}$ and as a wave plate at $\lambda_{2}$.For the general case when ququart is not a product state of two qubits the
transformation matrix $\hat{G}$:
\begin{equation}
|\Psi\rangle^{out}=\hat{G}|\Psi\rangle^{in} \label{eq:trans}
\end{equation}
can be found simply by making use of the Heizenberg picture. We use the equivalent
representation of (\ref{eq:state}) using creation and annihilation
operators:
\begin{equation}
|\Psi\rangle^{in}=[c_1a_1^{\dagger}a_2^{\dagger}+c_2a_1^{\dagger}b_2^{\dagger}+c_3b_1^{\dagger}a_2^{\dagger}+c_4b_1^{\dagger}b_2^{\dagger}]|vac\rangle.
\label{eq:operstate}
\end{equation}

SU(2) transformation between input $a_j$, $b_j$  and output
$a_j^{\prime}$, $b_j^{\prime}$ operators is given by:
\begin{equation}{{\begin{array}{*{20}c}
a_j^{\prime}=t_{j}a_j+r_{j}b_j \hfill \\
b_j^{\prime}=-r_{j}^{*}a_j+t_{j}^{*}b_j,\hfill \\
\end{array} }}
\label{eq:transform1}
\end{equation}
where

\begin{equation}
\begin{array}{cc}
 t_{j} = \cos \delta_{j} + i\sin \delta_{j} \cos 2\alpha_{j},\\
\quad r_{j} = i\sin \delta_{j} \sin 2\alpha_{j} , \delta_{j} =
{\pi (n_{o_{j}} - n_{e_{j}}) h/ \lambda_{j} }.
\end{array}
\end{equation}
Here $t_{j}$ and $r_{j}$ are the amplitude transmission and
reflection coefficients of the waveplate, $\delta_{j}$ is its
optical thickness, $h$ is the geometrical thickness, $\alpha_{j}$
is the orientation angle between the optical axis of the plate and
vertical direction. Lower indices $j=1,2$ numerate the frequency
modes of photons. Substituting (\ref{eq:transform1}) into
(\ref{eq:operstate}) one can immediately find that the unitary
transformation on state (\ref{eq:state}) is given by $4\times 4$
matrix which is obtained by a direct product of two $2\times 2$
matrices describing the transformation performed on each photon
\cite{dnk:97}:

\begin{equation}\footnotesize
\hat{G} \equiv \left( {{\begin{array}{*{20}c}
 {t_{1}t_{2}} \hfill & {t_{1}r_{2}} \hfill & {r_{1}t_{2}} \hfill & {r_{1}r_{2}} \hfill \\
 {-t_{1}r_{2}^*} \hfill & {t_{1}t_{2}^*} \hfill & {-r_{1}r_{2}^*} \hfill & {r_{1}t_{2}^*} \hfill \\
 {-r_{1}^*t_{2}} \hfill & {-r_{1}^*r_{2}} \hfill & {t_{1}^*t_{2}} \hfill & {t_{1}^*r_{2}} \hfill \\
 {r_{1}^*r_{2}^*} \hfill & {-r_{1}^*t_{2}^*} \hfill & {-t_{1}^*r_{2}^*} \hfill & {t_{1}^*t_{2}^*} \hfill \\
\end{array} }} \right) =
\left({{\begin{array}{*{20}c}
 {t_{1}} \hfill & {r_{1}} \hfill \\
 {-r_{1}^*} \hfill & {t_{1}^*} \hfill \\
\end{array} }} \right)\otimes
\left({{\begin{array}{*{20}c}
 {t_{2}} \hfill & {r_{2}} \hfill \\
 {-r_{2}^*} \hfill & {t_{2}^*} \hfill \\
\end{array} }} \right),
\label{eq:gmatrix}
\end{equation}

It is important to note that the optical thickness $\delta$ depends
on the wavelength, so parameters $t, r$ which determine the result
of transformation differ for photons forming biphoton-ququart. Thus
for the ququart transformation (without taking non-essential phase
term into account) considered above (\ref{eq:vv}) the matrix has the form

\begin{equation}
\hat{g}\equiv \left( {{\begin{array}{*{20}c}
 {0} \hfill & {0} \hfill & {1} \hfill & {0} \hfill \\
 {0} \hfill & {0} \hfill & {0} \hfill & {1} \hfill \\
 {1} \hfill & {0} \hfill & {0} \hfill & {0} \hfill \\
 {0} \hfill & {1} \hfill & {0} \hfill & {0} \hfill \\
 \end{array} }} \right).
 \label{eq:platematrix}
\end{equation}

The same dichroic plate performs unitary transformations between
polarization Bell states:
$g|\Phi^{\pm}\rangle=|\Psi^{\pm)}\rangle$, and
$g|\Psi^{\pm}\rangle=|\Phi^{\pm}\rangle$, where
$|\Phi^{\pm}\rangle$ and $|\Psi^{\pm}\rangle$ are represented by
ququarts with $c_{1}=\pm c_{4}=\frac{1}{\sqrt{2}}$ and $c_{2}=\pm
c_{3}=\frac{1}{\sqrt{2}}$ correspondingly. Similar transformations
with frequency non-degenerate biphotons propagating in single
spatial mode have been realized in \cite{freqBelltrans}.

\section{Experimental implementation} In this section we consider a several
methods of ququart state preparation and their measurements.

\subsection{Preparation} In general to prepare arbitrary ququart state
(\ref{eq:state}) it is necessary to use four nonlinear crystals
arranged in such a way that each crystal emits coherently one basic
state in the same direction. But in particular cases reduced set of
crystals is quite enough to generate specific ququart states which
can be used in practice. In experiments described below we used the
following methods to generate sub-set of ququarts.
\subsubsection{Separable states}
If nonlinear crystal generates one of the basic state
(\ref{eq:basis}) then applying the transformation
(\ref{eq:gmatrix}) one can get the sub-set of ququarts which are
the product states of pair polarization qubits (Fig.3).
\begin{figure}[!hb]
\includegraphics[width=0.28\textwidth]{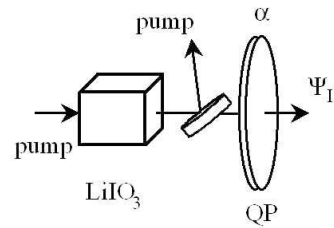}
\caption{Preparation of states $\Psi_{I}$. The nonlinear crystal
generates the basis ququart state(s) via type-I SPDC. Then the
state is transformed by the wave plate \textbf{QP} whose
parameters depend on its orientation angle $\alpha$.}
\end{figure}
We used the basis state $|V_{1}V_{2}\rangle$, so the states
generated in this way have the following structure:
\begin{equation}
|\Psi_{I}(\alpha)\rangle=\left( {{\begin{array}{*{20}c}
{r_{1}r_{2}} \hfill \\
{r_{1}t_{2}^*} \hfill \\
{t_{1}^*r_{2}} \hfill \\
{t_{1}^*t_{2}^*} \hfill \\
\end{array} }} \right).
\label{eq:gbasis}
\end{equation}
Since coefficients $r_{j}$ and $t_{j}$ depends on the the
orientation angle $\alpha_{j}$ there is a simple way to transform
the state by rotating the retardant plate. As we will show below, a
usage of a single crystal allows one to prepare the whole sub-set of
ququarts which can be used for quantum key distribution.
\subsubsection{Entangled states}
To prepare the ququarts with  $c_{1}c_{4}\neq c_{2}c_{3}$ it was
sufficient to use two crystals like it was done when frequency
non-degenerated Bell-states have been generated
\cite{freqBellprep} (Fig.4).

\begin{figure}[!hb]
\includegraphics[width=0.4\textwidth]{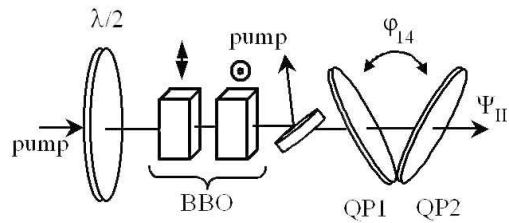}
\caption{Preparation of states $\Psi_{II}$. Two nonlinear crystals
generates pair of the basis ququart states via type-I SPDC. The
relative phase between states $\varphi_{14}$ is controled by two
quartz plates \textbf{QP1, QP2}. The state amplitudes are
controlled with the help of half-lambda plate.}
\end{figure}

In order to introduce a phase shift $\varphi_{14}$ between
horizontally and vertically polarized type-I biphotons two 1 mm
quartz plates (QP1,2) that can be rotated along the optical axis
were used. By tilting the plates the family of ququarts was
generated:
\begin{equation}
|\Psi_{II}(\phi)\rangle=\left( {{\begin{array}{*{20}c}
{|c_1|} \hfill \\
{0} \hfill \\
{0} \hfill \\
{|c_4|e^{-i\varphi_{14}}} \hfill \\
\end{array} }} \right),
 \label{eq:prepar1}
\end{equation}
where real amplitudes $|c_1|, |c_4|$ were controlled by the half
wave plate inserted into the linearly polarized pump beam. Some
remarkable interference phenomena with
$|\Psi_{II}(\phi)\rangle$-like states have been observed. For
example, when the phase of  $\lambda_1$-photon is varied, then
modulation with the same wavelength is observed in coincidence
rate, while changing the phase of $\lambda_2$-photon it is this
modulation that is manifested in coincidences \cite{freqBellprep}. Another
effect related to biphotons-ququarts that is widely discussed in
quantum optics is "hidden polarization". Polarization
transformations of frequency non-degenerate Bell-states
experimentally have been studied in \cite{freqBelltrans}.
\subsection{Measurement}  Basically it is necessary to perform $D^2$
independent measurements for complete reconstruction of the
arbitrary qudit density matrix. So for $D=4$ the total number of
measurements is equal to sixteen. Since the only way to measure
biphoton is to register a coincidence of photocounts, we chose a
Brown-Twiss scheme supplied with polarization elements to perform
projective measurements. A coincidence click that occurs when
signals from two detectors fall into the coincidence window is
considered a registered event. Since the coincidence click appears
at the output with some probability, a statistical treatment of the
outcomes becomes very important. Another point that we would like to
notice is that according to Bohr's complementarity principle, there
is no way to measure all moments (\ref{eq:diag}, \ref{eq:nondiag})
at the same time, dealing with a single ququart only. So to perform
a complete set of measurements one needs to generate a lot of
representatives of a quantum ensemble.
\par In order to measure the ququarts we developed two protocols.
\subsubsection{Protocol 1.}
 The idea of the first method is to split the
ququart state $| {\Psi_{in} } \rangle $ into two frequency modes and
to perform transformations independently on each of the photon from
a pair (Fig. 5). Usage of a dichroic beamsplitter allows one to
separate frequency modes in space and reduces this protocol to that
one developed in \cite{Kwiat:01} for two spatially separated
polarization qubits. We called this method as "frequency selective"
because frequency mode separation is basically applied to ququart
before subjecting it to polarization projective measurements.
\begin{figure}[!ht]
\includegraphics[width=0.4\textwidth]{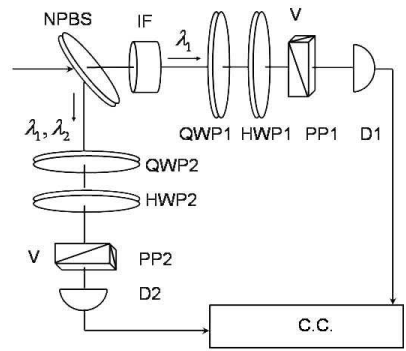}
\caption{Measurement block for Protocol 1. The Brown-Twiss scheme
for measuring intensity correlation between two polarization
modes. After spatial separation at the non-polarizing beam
splitter (\textbf{NPBS}), signal $(\lambda_2)$ and idler
$(\lambda_1)$ photons propagate through the quarter-
(\textbf{QWP1,2}) and half (\textbf{HWP1,2}) wave plates ,
polarizing prisms (\textbf{PP1,2}) in two channels. The
interference filter (\textbf{IF}) placed in the upper arm and
centered at $\lambda_1$ selects frequencies modes of ququart.
Finally, photons are registered by detectors (\textbf{D1,2}). The
coincidence rate from the output of the coincidence circuit
(\textbf{CC}) is proportional to the fourth moment in the field
$\langle {R} \rangle $.}
\end{figure}
The polarization transformations can be done using polarization
filters placed in front of detectors. Each filter consists of a
sequence of quarter- and half-wave plates and a polarization prism,
which picks out vertical polarization. In our experiments we kept
the protocol developed in \cite{Kwiat:01} unchanged, although there
are many other configurations of polarization filters leading to the
matrix $K_{4}$ reconstruction. Instead of dichroic beamsplitter we
used a (50/50)\% polarization insensitive beamsplitter and a
narrowband filter centered at $\lambda_{1}$ was put in front of the
first detector. In such a way we postselected the desirable events
in registering coincidences. Approximately in quarter of trials, the
photons with central wavelength $\lambda_{1}$ forming a biphoton are
going to the first detector, while the conjugated ones (with central
wavelength $\lambda_{2}$) are going to the second detector. In the
remaining cases, the events are not registered because they either
do not contribute to coincidences or corresponding photons do not
pass through the interference filter.
\par In the Heisenberg representation the polarization
transformation for each dichroic beam-splitter output port is
given by:
\begin{eqnarray}
 \left( {{\begin{array}{*{20}c}
 {{a_{j}}'^\dagger} \hfill \\
 {{b_{j}}'^ \dagger} \hfill \\
\end{array} }} \right) = \left( {{\begin{array}{*{20}c}
 0 \hfill & 0 \hfill \\
 0 \hfill & 1 \hfill \\
\end{array} }} \right)
D_{\lambda_{j} /2} ( {\delta_{j} = \pi/2,\theta }) \nonumber\\
\times D_{\lambda_{j}/4} ( {\delta_{j} = \pi /4,\chi })
 \left( {{\begin{array}{*{20}c}
 {\frac{1}{\sqrt 2} } \hfill & 0 \hfill \\
 0 \hfill & {\frac{1}{\sqrt 2} } \hfill \\
\end{array} }} \right)
\left( {{\begin{array}{*{20}c}
 {a_{j}^\dagger} \hfill \\
 {b_{j}^ \dagger} \hfill \\
\end{array} }} \right).
 \label{eq:transform}
\end{eqnarray}
We designate with $j=1, 2$ different output ports of beamsplitter
which correspond to different wavelengths $\lambda_{1,2}$. Four
$2\times2$ matrixes in the right-hand side of (\ref{eq:transform})
describe the action of the non-polarizing beam-splitter,
$\lambda/4$-, $\lambda/2$- plates and vertical polarization prism
on the state vector of each photon;
\begin{equation}
D_{\lambda_{j}/2,\lambda_{j}/4} = \left( {{\begin{array}{*{20}c}
 t_{j} \hfill & r_{j} \hfill \\
 { - r_{j}^\ast } \hfill & {t_{j}^\ast } \hfill \\
\end{array} }} \right),
\nonumber\end{equation} where $r_{j}$ and $t_{j}$ - are the
coefficients introduced in Eq. (23), so for a $\lambda_{j}/4$-plate
($\delta_{j} = \pi/4$),

\begin{subequations}
\begin{equation}
t_{\lambda_{j}/4} = \frac{1}{\sqrt 2}({1 + i\cos 2\chi_{j} }),
r_{\lambda_{j}/4} = \frac{i}{\sqrt 2 }\sin 2\chi_{j}
\end{equation}
and for a
 $\lambda_{j}/2$-plate ($\delta_{j} = \pi/2$)

\begin{equation}
t_{\lambda_{j}/2} = i\cos ( {2\theta_{j} } ), r_{\lambda_{j}/2} = i\sin
( {2\theta_{j} } ). \end{equation} \end{subequations}Thus, there are
four real parameters (two for each channel) that determine
polarization transformations. Namely, these parameters are
orientation angles for two pairs of wave plates: $\theta _1 , \chi
_1 , \theta _2 , \chi _2 .$ Also we would like to notice that there
is another parameter that affects strongly on result of polarization
transformations. It is a wavelength(s) of down-converted photon(s).
For example varying the frequency spectrum of SPDC by tilting a
crystal which generates biphotons one can significantly change the
transformed state (\ref{eq:trans}) using fixed parameters of the
wave plates such as geometrical thickness $h$ and its orientation
$\alpha$. This property makes biphoton-based ququarts much more
flexible to be controlled than biphoton-qutrits and allows one to
choose convenient regimes for operations with them.

As it was mentioned above, the output of the Brown-Twiss scheme is
the coincidence rate of the pulses coming from two detectors
$D_{1}$ and $D_{2}$. The corresponding moment of the fourth order
in the field has the following structure:

\begin{equation} R_{1,2} \propto \langle{
{{b'}_1}^\dagger{{b'}_2}^\dagger{b'}_1{b'}_2}\rangle = R( {\theta
_1 ,\chi _1 , \theta _2, \chi _2 } )
\label{eq:selective}
\end{equation}
 In the most
general case this moment contains a linear combination of ten
moments (\ref{eq:diag}, \ref{eq:nondiag}) forming the matrix $K_4$.
So the main purpose of the state reconstruction procedure is
extracting these moments from the set-up outcomes by varying the
four parameters of the polarization Brown-Twiss scheme. Consider
some special examples, which give the corresponding lines in the
complete protocol introduced below (Table I). The measurement of
first four moments (\ref{eq:diag}) is trivial procedure. For
instance the third line in the (Table I) corresponds to selection
the basis state $|V_{1}V_{2}\rangle$. The other three upper lines
correspond to the measurement of other basis states
$|H_{1}H_{2}\rangle$, $|H_{1}V_{2}\rangle$, and
$|V_{1}H_{2}\rangle$. Remaining lines of protocol show how to
measure one of the complex moments (\ref{eq:nondiag}). To measure
the real part of the moment $E$, let us set the wave-plates in the
Brown-Twiss scheme in the following way.

\noindent  The first channel:

\begin{subequations}
\begin{equation}
\lambda /4: \chi _1=0^\circ, D_{\lambda /4} = \frac{1}{\sqrt 2
}\left( {{\begin{array}{*{20}c}
 1+i \hfill & 0 \hfill \\
 0 \hfill & 1-i \hfill \\
\end{array} }} \right);
\end{equation}
\begin{equation}
\lambda/2: \theta _1= 45^\circ, D_{\lambda/2} = \left(
{{\begin{array}{*{20}c}
 0 \hfill & i \hfill \\
 i \hfill & 0 \hfill \\
\end{array} }} \right).
\end{equation}
\end{subequations}
The second channel:
\begin{subequations}
\begin{equation}
\lambda/4 :\chi _2=-45^\circ, D_{\lambda/4} = \frac{1}{\sqrt 2
}\left( {{\begin{array}{*{20}c}
 {1} \hfill & -i \hfill \\
 -i \hfill & {1} \hfill \\
\end{array} }} \right);
\end{equation}
\begin{equation}
\lambda /2: \theta_2= -22.5^\circ, D_{\lambda/2} = \frac{1}{\sqrt 2}\left(
{{\begin{array}{*{20}c}
 i \hfill & -i \hfill \\
 -i \hfill & -i \hfill \\
\end{array} }} \right).
\end{equation}
\end{subequations}

 Substituting these matrices into Eq. (28) and taking
into account the commutation rules for the creation and
annihilation operators it is easy to get the final moment to be
measured:

\begin{equation}
R = \langle \Psi |b_1^\dagger b_2^\dagger b_1 b_2 | \Psi \rangle =
1/8( {A + C + 2\textrm{Re}E}). \nonumber\end{equation} A complete
set of the measurements called the tomography protocol is presented
in Table I. Each row corresponds to the setting of the plates to
measure the moment highlighted in the sixth column. The last column
corresponds to amplitude of the process (see below).
\begin{table*}
\caption{Protocol 1. Each line contains orientation of the half
($\theta_{s,i}$) and quarter ($\chi_{s,i}$) wave plates in the
measurement block. Last two columns show the corresponding moment
$R_\nu$ and the process amplitude $M_\nu(\nu =1,..16)$. Sign "-" in
orientation of plates for reflection channel is introduced to
compensate $\pi$-phase shift occurring by reflectance from the
beamsplitter.}

\begin{tabular}{|c|c|c|c|c|c|c|c|} \hline
 & \multicolumn{4}{c|}{Set-up parameters}
 & Moments to be measured & Amplitude of the
process\\\hline
  $\nu$&$\chi_s$&$\theta_s$&$\chi_i$&$\theta_i$&$R_\nu=|M_\nu|^2$&$M_\nu$\\\hline
  1. & 0 & $45^\circ$ & 0 & $-45^\circ$ & $A/4$ & $c_{1}/2$\\\hline
  2. & 0 & $45^\circ$ & 0 & 0 & $B/4$ & $c_{2}/2$\\\hline
  3. & 0 & 0 & 0 & 0 & $D/4$ &$c_{4}/2$\\\hline
  4. & 0 & 0 & 0 & $-45^\circ$ & $C/4$ & $c_{3}/2$\\\hline
  5. & 0 & $22.5^\circ$ & 0 & $-45^\circ$ & $1/8(A+C+2\textrm{Im}F)$ & $\frac{1}{2\sqrt 2
}(c_{1}-ic_{3})$\\\hline
  6. & 0 & $22.5^\circ$ & 0 & 0 & $1/8(B+D+2\textrm{Im}K)$ & $\frac{1}{2\sqrt 2
}(c_{2}-ic_{4})$ \\\hline
  7. & $45^\circ$ & $22.5^\circ$ & 0 & 0 & $1/8(B+D-2\textrm{Re}K)$& $\frac{1}{2\sqrt 2
}(c_{2}-c_{4})$ \\\hline
  8. & $45^\circ$ & $22.5^\circ$ & 0 & $-45^\circ$ & $1/8(A+C-2\textrm{Re}F)$&  $\frac{1}{2\sqrt 2
}(c_{1}-c_{3})$\\\hline
  9. & $45^\circ$ & $22.5^\circ$ & 0 & $-22.5^\circ$ & $1/16(A+B+C+D)-1/8(\textrm{Im}E+\textrm{Re}F-\textrm{Im}G+\textrm{Im}I+\textrm{Re}K+\textrm{Im}L)$ & $\frac{1}{4}[(c_{1}-c_{3})+i(c_{2}-c_{4})]$ \\\hline
  10. & $45^\circ$ & $22.5^\circ$ & $-45^\circ$ & $-22.5^\circ$ & $1/16(A+B+C+D)-1/8(\textrm{Re}F-\textrm{Re}E+\textrm{Re}G+\textrm{Re}I+\textrm{Re}K-\textrm{Re}L)$ &$\frac{1}{4}[c_{1}+c_{2}-c_{3}-c_{4}]$\\\hline
  11. & 0 & $22.5^\circ$ & $-45^\circ$ & $-22.5^\circ$ & $1/16(A+B+C+D)+1/8(\textrm{Im}F+\textrm{Re}E+\textrm{Im}G+\textrm{Im}I+\textrm{Re}L+\textrm{Im}K)$ &$\frac{1}{4}[(c_{1}+c_{2})-i(c_{3}+c_{4})]$\\\hline
  12. & 0 & $45^\circ$ & $-45^\circ$ & $-22.5^\circ$ & $1/8(A+B+2\textrm{Re}E)$ & $\frac{1}{2\sqrt 2
}(c_{1}+c_{2})$\\\hline
  13. & 0 & 0 & $-45^\circ$ & $-22.5^\circ$ & $1/8(C+D+2\textrm{Re}L)$ & $\frac{1}{2\sqrt 2
}(c_{3}+c_{4})$\\\hline
  14. & 0 & 0 & $-90^\circ$ & $-22.5^\circ$ & $1/8(C+D+2\textrm{Im}L)$ & $\frac{1}{2\sqrt 2
}(c_{3}-ic_{4})$\\\hline
  15. & 0 & $45^\circ$ & $-90^\circ$ & $-22.5^\circ$ & $1/8(A+B+2\textrm{Im}E)$ & $\frac{1}{2\sqrt 2
}(c_{1}-ic_{2})$\\\hline
  16. & 0 & $22.5^\circ$ & $-90^\circ$ & $-22.5^\circ$ & $1/16(A+B+C+D)+1/8(\textrm{Im}F+\textrm{Im}E-\textrm{Re}G+\textrm{Re}I+\textrm{Im}L+\textrm{Im}K)$ & $\frac{1}{4}[(c_{1}-c_{4})-i(c_{2}+c_{3})]$\\\hline

\end{tabular}

\end{table*}
\par Amplitude of the quantum process, corresponding to passing
down-converted photons through the measurement set-up for each
configuration of wave plates is described by the following equation:
\begin{equation}
M_{\nu}(\delta_{1},\theta_{k},
\delta_{2},\varphi_{l})=\frac{1}{2}[a_1a_2c_1
+a_1b_2c_2+b_1a_2c_3+b_1b_2c_4].
 \label{eq:amplitude}
\end{equation}
where
\begin{equation}\label{eq:coeficient}
\begin{array}{cc}
a_1=-{r_{\lambda/2}}(\theta_1)t^*_{\lambda/4}(\chi_1)-{t_{\lambda/2}}(\theta_1)r_{\lambda/4}(\chi_1),
&\\
a_2=-{r_{\lambda/2}}(\theta_2)t^*_{\lambda/4}(\chi_2)-{t_{\lambda/2}}(\theta_2)r_{\lambda/4}(\chi_2), &\\
b_1=-{r_{\lambda/2}}(\theta_1)r^*_{\lambda/4}(\chi_1)+{t_{\lambda/2}}(\theta_1)t_{\lambda/4}(\chi_1), &\\
b_2=-{r_{\lambda/2}}(\theta_2)r^*_{\lambda/4}(\chi_2)+{t_{\lambda/2}}(\theta_2)t_{\lambda/4}(\chi_2).
\end{array}
\end{equation}

The complete reconstruction of the input state $|\Psi^{in}\rangle$
was performed according to the algorithm considered in our previous
work \cite{ourPRA:04}.
\par Concluding this section we would like to stress that if usual
frequency insensitive (broadband) beamsplitter is used and no
interference filters are inserted either in front of or behind this
beamsplitter, then "non-selective" method of ququart measurement is
possible. In this case one does not need to use the interference
filter to select the wavelength in each channel of Brown-Twiss
scheme. In other words each detector is allowed to register photons
with different frequencies and the fourth moment in the field to be
measured becomes
\begin{equation}
R_{1,2} \propto \langle{
({{b'}_{1}+{b'}_2})^\dagger_{Detector2}({b'}_{1}+{b'}_2})_{Detector1}\rangle
\label{eq:nonselective}
\end{equation}
instead of (\ref{eq:selective}). In non-selective method the expression connecting observable value $R_{1,2}$ with components of coherence matrix (\ref{eq:diag},
\ref{eq:nondiag}) is bulky and it is not reasonable to utilize it for the state reconstruction.
 However in particular case we applied
non-selective method to ququart reconstruction procedure and
developed the special protocol, which is introduced in the next
section.
\subsubsection{Protocol 2.}
In the second protocol the whole ququart state (\ref{eq:stategen})
is subjected to the linear transformations using a set of the
retardant plates \cite{JETPLett}. Then projective measurements were
applied to the state $|\Psi^{in}\rangle$ after it being transformed
by the plates. Using the retardant plates with fixed optical
thickness, one can reconstruct the state, varying the orientation of
the plates (Fig. 6).
\begin{figure}[!ht]
\includegraphics[width=0.38\textwidth]{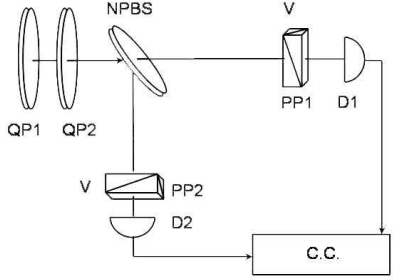}
\caption{Measurement block for Protocol 2. After the input state
is subjected to transformations by wave plates (\textbf{QP1,
QP2}), the projective measurements selecting vertical
polarizations of ququart components were performed.}
\end{figure}
Mathematically one needs to derive an equations set connecting
real and imaginary parts of moments (\ref{eq:diag},
\ref{eq:nondiag}) with coincidence rate. It turns out that the
complete set of moments (\ref{eq:diag}, \ref{eq:nondiag}) can be
extracted from the projective measurements if two different plates
were used.
\begin{equation}
|\Psi^{out}\rangle_{kl}=\hat{G}(\delta_{1},\theta_{k})\hat{G}(\delta_{2},\varphi_{l})|\Psi^{in}\rangle,
 \label{eq:protocol1}
\end{equation}
where $\theta_{k}$ and $\varphi_{l}$ are the orientation angles of
the first and second plates. The parameters of the plates, i.e.
optical thicknesses for different wavelengths
$\delta_{\lambda{j}}^{(1,2)}$ and their orientations are supposed
to be known. The projective measurements were realized by means of
pair of polarization prisms, transmitting vertical polarization
and settled in front of each single photon detector. So the number
of coincidences is
\begin{equation}
R_{kl}\propto|\langle V_{1}V_{2}|\Psi_{out}\rangle_{kl}|^2.
 \label{eq:coinc1}
\end{equation}
\par In order to reconstruct the state it is sufficient to
perform at least four measurements corresponding to different
orientations of the second plate $\varphi_{l}$ and repeat this
procedure at least four times varying orientation of the first plate
$\theta_{k}$. But in experiments we performed redundant number of
measurements to increase the accuracy and finally used 36
orientation of the second plate. Thus total number of measurements
$\mu$ in this protocol was equal to $4\times36=148$. Therefore the
Protocol 2 includes 148 lines instead of 16 lines used in Protocol
1.

\subsection{Experimental setup.} For generation of biphoton-based
ququarts we used lithium-iodate 15 mm crystal pumped with 5 mW cw-
helium-cadmium laser operating at 325 nm. The orientation of the
crystal is chosen at $58.98^\circ$ with respect to the pump wave
vector that the down-converted photons at $\lambda_{1}=702 nm$ and
$\lambda_{2}=605 nm$ had been generated. For some particular cases
we selected biphoton-ququarts at wavelengths $\lambda_{1}=667 nm$
and $\lambda_{2}=635 nm$. Thus the either $|V_{702}V_{605}\rangle$
or $|V_{667}V_{635}\rangle$ states were used as an initial states.
Then, this state was subjected to transformations done by dichroic
waveplates to prepare the sub-set of ququarts with $c_{1}c_{4}=
c_{2}c_{3}$ (\ref{eq:gbasis}). That sub-set of states was used to be
reconstructed. In particular we used the 0.988-mm (or 0.315-mm ) length quartz
plate and changed it orientation $\alpha$, which served as a
parameter. Since the thickness of the plate, quartz dispersion and
orientation $\alpha$ are supposed to be known, we were able to
calculate the result of the state transformation with high accuracy.
When Protocol 2 was applied for ququart reconstraction we used the
quartz plates with thicknesses $0.821 mm$ (QP1) and $0.715 mm$
(QP2). Both protocols were applied to reconstruct the initial state
$|\Psi_{I}(\alpha)\rangle$. In the case when sub-set $c_{1}c_{4}\neq
c_{2}c_{3}$ had been generated we used two consecutive 1.8 mm thick
type-I BBO (beta-barium borate) crystals whose optical axis are
oriented perpendicularly with respect to each other at $36.33^\circ$
with respect to the pump wave vector.
 An interference filter centered either at
$702 nm$ or at $635 nm$ and with a FWHM bandwidth of $3nm$ was
placed in transmitted arm to make a spectral selection of one
photon of a pair, while the photon with conjugated frequency was
sent to a reflected arm. As it was already mentioned above this
scheme is equivalent to that one used in \cite{Kwiat:01}. The only
difference between our scheme and developed in \cite{Kwiat:01}
resulted from operating with frequency non-degenerate collinear
regime of SPDC instead of non-collinear, frequency degenerate
regime used in \cite{Kwiat:01}. Without loss of generality, this
scheme also allows to reconstruct any arbitrary polarization
ququart state by registering coincidence rate for different
projections that are done by the polarization filters located in
each arm. Each filter consists of a zero-order quarter- and
half-waveplate and a fixed analyzer. Two Si-APD's linked to a
coincidence scheme with 1.5 nsec time window were used as single
photon detectors.

\subsection{Results and discussion}
We applied both protocols to measure the set of states $\Psi_{I}$,
generated with $|V_{667nm}V_{635nm}\rangle$ transformed by quartz
plate QP, like it is shown in Fig. 3. We have tested the states
which were generated for two orientation angles of plate QP $\alpha=
0^\circ, 90^\circ$. According to the Protocol 2 four sets of
measurements was performed for each input state, so totally we
performed 148 measurements of coincidence rate as a function of
orientation angles $\varphi_{l}$ of QP1 and $\theta_{k}$ of QP2. The
dependence of coincidence rate on the orientation $\varphi_{l}$ for
$\theta_{k}= 90^\circ (a)$, $105^\circ (b)$, $120^\circ (c)$, and
$135^\circ (d)$ is shown in Fig. 7 for the input state corresponding
to orientation $\alpha = 20^\circ$ of the plate forming the state.
\begin{figure}[!ht]
\includegraphics[width=0.5\textwidth]{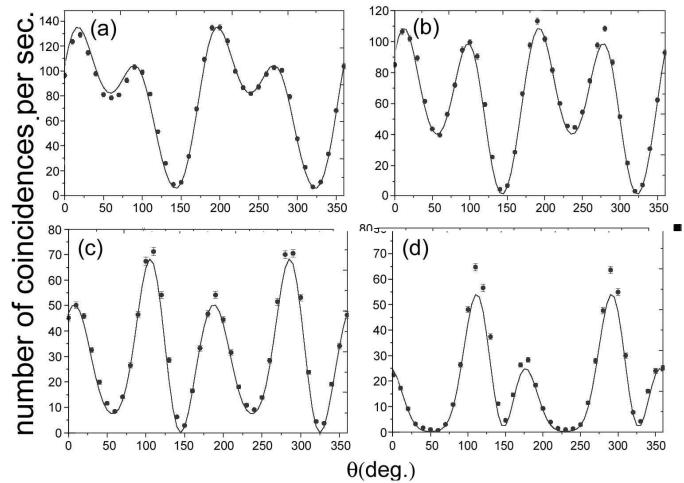}
\caption{Dependence of coincidence rate on tilting angle
${\theta}$}
\end{figure}

The solid curve gives the theoretical behavior of the normalized
fourth moment, the dots with bars show the experimental data. The
tables II-IV show the results of statistical reconstruction of
ququarts $|\Psi_{I}\rangle$ performed with root estimation method.
The tables show theoretical and experimental four-dimensional
state-vectors $c_{theory}$, $c_{exp}$ as well as fidelity $F$
defined by
\begin{equation}
F = | \langle c_{theory}| c_{exp}\rangle|^2 .
\end{equation}
  In table II we collected the data
measured by Protocol 1 (ququarts with $\lambda_{1}=702 nm$ and
$\lambda_{2}=605 nm$, $h$(QP)=0.315mm). Table III corresponds to measurements
performed by Protocol 2 (ququarts with $\lambda_{1}=667 nm$ and
$\lambda_{2}=635 nm$, $h$(QP)=0.988mm). And finally we demonstrate the validity of
preparation, transformation, and measurement procedures performed
over biphotons-ququarts by reconstruction of the Bell states
$|\Phi^{\pm}\rangle$ ($\lambda_{1}=702 nm$ and $\lambda_{2}=605
nm$), measured along with Protocol 1 and shown in Table IV. As it is
seen from the tables the accuracy of state reconstruction is a
little bit lower when Protocol 1 has been applied. It results from
the fact that Protocol 1 exploits minimal number of measurements
(16) which is needed for four-state system reconstruction. At the
same time since redundant number of measurements (148) was involved
when applying the second protocol, the highest accuracy was achieved
like in the case of biphotons-qutrits \cite{ourPRA:04}. The
respectively low fidelity for Bell states reconstruction (Table IV)
is explained by experimental imperfections at the preparation stage.
Preliminary estimation of the prepared state quality can be
extracted from visibility displayed during interference experiments
with these states. For example, typical visibility meaning revealed
when space-time interference observed was $0.9$ \cite{interference}.
Another factor limiting the accuracy of the state reconstruction is
the finite number of events to be registered. In the case of
$|\Psi_{I}\rangle$-set of states we collected about several
thousands of coincidences during accumulation time whereas in the
case of $|\Psi_{II}\rangle$-set  only a few hundreds coincidences
were registered in total.

\begin{table}[!ht]
\caption{}
\begin{tabular}{|c|c|c|c|}
\hline \multicolumn{4}{|c|}{protocol 1} \\ \hline $\alpha$ (deg.) &
theory & experiment & $F$ \\ \hline 0 &
\begin{tabular}{c}
0 \\
0 \\
0 \\
1
\end{tabular}
&
\begin{tabular}{c}
-0.0295 - 0.0306i \\
0.0543 - 0.0202i \\
-0.0154 - 0.0093i \\
0.9972
\end{tabular}
& 0.995 \\ \hline 10 &
\begin{tabular}{c}
-0.0015 - 0.0229i \\
0.0038 - 0.0050i \\
0.2013 + 0.1735i \\
0.9638
\end{tabular}
&
\begin{tabular}{c}
0.0584 - 0.1926i \\
0.0073 + 0.0170i \\
0.1633 + 0.1233i \\
0.9577
\end{tabular}
& 0.963 \\ \hline 20 &
\begin{tabular}{c}
-0.0021 - 0.0386i \\
0.0154 - 0.0162i \\
0.3430 + 0.3625i \\
0.8654
\end{tabular}
&
\begin{tabular}{c}
0.0015 - 0.0326i \\
0.0019 - 0.0660i \\
0.3967 + 0.2085i \\
0.8909
\end{tabular}
& 0.976 \\ \hline 30 &
\begin{tabular}{c}
-0.0015 - 0.0445i \\
0.0337 - 0.0225i \\
0.3530 + 0.5716i \\
0.7383
\end{tabular}
&
\begin{tabular}{c}
-0.0310 - 0.0542i \\
0.0166 - 0.0416i \\
0.4526 + 0.5386i \\
0.7065
\end{tabular}
& 0.991 \\ \hline 40 &
\begin{tabular}{c}
-0.0005 - 0.0440i \\
0.0511 - 0.0116i \\
0.1601 + 0.7466i \\
0.6421
\end{tabular}
&
\begin{tabular}{c}
0.0112 - 0.0823i \\
0.1174 - 0.0328i \\
0.1087 + 0.8363i \\
0.5167
\end{tabular}
& 0.970 \\ \hline
\end{tabular}
\end{table}

\begin{table}[!ht]
\caption{ }
\begin{tabular}{|c|c|c|c|}
\hline \multicolumn{4}{|c|}{protocol 2} \\ \hline $\alpha$ (deg.) &
theory& experiment& $F$ \\ \hline 0 &
\begin{tabular}{c}
0 \\
0 \\
0 \\
1
\end{tabular}
&
\begin{tabular}{c}
-0.0555-0.0204i \\
-0.0059+0.005i \\
-0.0425+0.0052i \\
0.9973
\end{tabular}
& 0.996 \\ \hline 20 &
\begin{tabular}{c}
0.8097 \\
-0.4568-0.3527i \\
-0.0103-0.0859i \\
-0.0316+0.0529i
\end{tabular}
&
\begin{tabular}{c}
0.8067 \\
-0.4847-0.3304i \\
0.023-0.0554i \\
-0.0174+0.0413i
\end{tabular}
& 0.998 \\ \hline
\end{tabular}
\end{table}

\begin{table}[!ht]
\caption{ }
\begin{tabular}{|c|c|c|}
\hline \multicolumn{3}{|c|}{protocol 1} \\ \hline theory&
experiment& $F$ \\ \hline
\begin{tabular}{c}
$0.707$ \\
0 \\
0 \\
$0.707$%
\end{tabular}
&
\begin{tabular}{c}
0.7326 \\
0.0818-0.0963i \\
0.0003-0.0281i \\
0.6131+0.2657i
\end{tabular}
& 0.941 \\ \hline
\begin{tabular}{c}
$0.707$ \\
0 \\
0 \\
-$0.707$%
\end{tabular}
&
\begin{tabular}{c}
0.6597 \\
0.2518+0.4692i \\
0.0897-0.0319i \\
-0.6155+0.3261i
\end{tabular}
& 0.934 \\ \hline
\end{tabular}
\end{table}

\section{ Polarization ququarts in QKD protocol}

\par The complete QKD protocol with four-dimensional polarization states exploits five mutually unbiased bases with four states in each \cite{mutual}. In terms of
biphoton states the first three bases consist of product
polarization states of two photons and last two bases consist of
two-photon entangled states:

\begin{equation}
\begin{array}{cc}
I.\quad|H_{1}H_{2}\rangle; \quad |H_{1}V_{2}\rangle;
\quad|V_{1}H_{2}\rangle; \quad|V_{1}V_{2}\rangle,\\
II. \quad|D_{1}D_{2}\rangle;
\quad|D_{1}\overline{D_{2}}\rangle;\quad
|\overline{D_{1}}D_{2}\rangle; \quad|\overline{D_{1}}\overline{D_{2}}\rangle, \\
III. \quad|R_{1}R_{2}\rangle; \quad|R_{1}L_{2}\rangle;\quad
|L_{1}R_{2}\rangle; \quad|L_{1}L_{2}\rangle,\\
IV. \quad|R_{1}H_{2}\rangle + |L_{1}V_{2}\rangle;\quad
|R_{1}H_{2}\rangle - |L_{1}V_{2}\rangle;\\
 |L_{1}H_{2}\rangle + |R_{1}V_{2}\rangle;\quad
|L_{1}H_{2}\rangle - |R_{1}V_{2}\rangle, \\
V. \quad|H_{1}R_{2}\rangle + |V_{1}L_{2}\rangle;\quad
|H_{1}R_{2}\rangle - |V_{1}L_{2}\rangle; \\ |H_{1}L_{2}\rangle +
|V_{1}R_{2}\rangle;\quad
 |H_{1}L_{2}\rangle - |V_{1}R_{2}\rangle.
 \end{array}
 \label{qkd:states}
\end{equation}

Here $|H\rangle\equiv|1\rangle$, $|V\rangle\equiv|0\rangle$,
$|D\rangle\equiv\frac{1}{\sqrt 2 }(|1\rangle+|0\rangle)$,
$\overline{D}\equiv\frac{1}{\sqrt 2 }(|1\rangle-|0\rangle)$,
$|R\rangle\equiv\frac{1}{\sqrt 2 }(|1\rangle+i|0\rangle)$,
$|L\rangle\equiv\frac{1}{\sqrt 2 }(|1\rangle-i|0\rangle)$ indicate
horizontal, vertical, +45 linear, -45 linear, right- and
left-circular polarization modes correspondingly, and lower indices
numerate the frequency modes of two photons. It has been proved
\cite{Tittel} that it is sufficient enough to use only first two or
three bases for the efficient QKD. Using incomplete set of bases one
sacrifices the security but enhances the key generation rate. Since
the realization of Bell measurements for the last two bases requires
a big experimental effort on both preparation and measurement stages
of a protocol, we will restrict ourselves to first three bases. As
we will show experimentally, all states from the first three bases
can be prepared with the help of a single non-linear crystal and
local unitary transformations. This is fundamental distinction with
respect to biphotons-qutrits where SU(2) transformations between
states from mutually unbiased bases are prohibited. We also present
a measurement scheme that allows to discriminate the states
belonging to one basis deterministically thus allowing its
implementation in realization of QKD protocol with polarization
ququarts.
\subsection{Experimental procedure.} Experimental setup for
generation of ququart states belonging first three bases
(\ref{qkd:states}) is shown in Fig. 8. To varify the state
generated in the set-up either protocol 1 or protocol 2 might be
used. Usually we applied protocol 1 for checking the ququart
state.
\begin{figure}[!ht]
\includegraphics[width=0.4\textwidth]{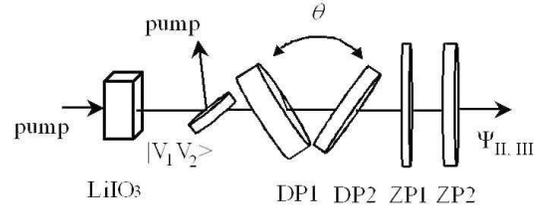}
\caption{Setup for preparation and measurement of ququarts which
can be used in QKD. Wave plates \textbf{DP1, DP2} oriented at
$45^\circ$ degrees with respect to vertical axis serve as dichroic
retardation plate with variable optical thickness which is
controlled by tilting angle $\theta$. Two zero-order plates
\textbf{ZP1, ZP2} allow to chose the basis.} \label{setup1}
\end{figure}
Let us consider as an example, the preparation of a state
$|H_{1}V_{2}\rangle$ from the initial state $|V_{1}V_{2}\rangle$.
This transformation can be achieved by a dichroic waveplate that
introduces a phase shift of $2\pi$ for a vertically polarized
photon at $605 nm$, a phase shift of $\pi$ for the conjugate
photon and that is oriented at $45^\circ$ to the vertical
direction. Using quarts plates as retardation material it is easy
to calculate that the thickness of the waveplate that does this
transformation should be equal to $3.406 mm$ \cite{order}. Since
these waveplates were not readily available and the result of
transformation is extremely sensitive to the small variations of
thickness, we used the following method to achieve the desired
thickness. Two quartz plates with geometric thicknesses of $3.716
mm$ (DP1) and $0.315 mm$ (DP2) with orthogonally oriented optical
axis were placed consecutively in the biphoton beam. The
consecutive action of these two waveplates correspond to the
action of quartz waveplate with an effective thickness of $3.401
mm$. If then one can tilt these waveplates towards each other by a
finite angle $\theta$, then the optical thickness of the effective
waveplate, formed by DP1 and DP2 will be changing, and, at a
certain value of $\theta$, the desired transformation will be
achieved. This corresponds to maximal coincidence rate when the
measurement part (protocol 1) is tuned to select a state
$|H_{1}V_{2}\rangle$. Monitoring the coincidences, one can obtain
the value of ${\theta}$ for which the maximum occurs. Then, fixing
the tilting angle at this value, one can perform a complete
quantum state tomography protocol in order to verify if the state
really coincides with the ideal. In order to change the basis from
$I$ to $II (III)$, zero order half- (quarter) waveplates $ZP1$
($ZP2$) oriented at $22.5^{\circ} (45^{\circ})$ were used. This
procedure is repeated for generation of any of the states. Fig. 9.
\begin{figure}[!ht]
\includegraphics[width=0.28\textwidth]{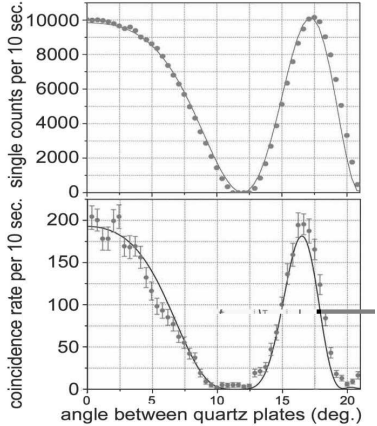}
\caption{Dependence of single counts (upper) and coincidences
(lower) on tilting angle ${\theta}$} \label{graphic}
\end{figure}
shows the coincidences and single count rates versus the change of
tilting angle $\theta$ which determines optical thickness of the
effective waveplate. If the measurement setup is tuned to select
the state $|H_{1}V_{2}\rangle$ then dependence of coincidence rate
on the plate optical thicknesses $\delta_{i}$ is given by formula:
\begin{equation}
R_{coin}\propto\langle{{a_1}^\dagger{b_2}^\dagger{a}_1{b}_2}\rangle
=\sin^2(\delta_{1})\cos^2(\delta_{2}),
\end{equation}
whereas the single counts distribution in the upper channel is
given by
\begin{equation}
I_{702nm}\propto\langle{{a_1}^\dagger{a}_1}\rangle=\sin^2(\delta_{1}),
\label{singles}
\end{equation}
where optical thickness depends on tilting angle as follows:
\begin{equation}
\delta_{j}=\frac{\pi h}{\lambda_{j}}[
\frac{(n_{e_{j}})^2}{\sqrt{(n_{e_{j}})^2-sin^{2}\theta}}-\frac{(n_{o_{j}})^2}{\sqrt{(n_{o_{j}})^2-sin^{2}\theta}}].
\label{optthickness}
\end{equation}
The solid lines in Fig. 9 show the theoretical curves. We performed
tomography measurements for the both maxima as well as for the
minimum. The minimum in coincidences occurs when intensity in any
channel drops to zero, so it is not a necessary condition for
distinguishing the orthogonal state to that one selected by given
settings of polarization filters. Nevertheless according to
calculations and our measurements the minimum in the coincidences at
Fig. 9 exactly refers to the state $|V_{1}H_{2}\rangle$. Starting
from the $|V_{702nm}V_{605nm}\rangle$, we also prepared and measured
the whole set of states from (\ref{qkd:states}) belonging to first
three bases. The result of state reconstruction is presented in
Table V.
\begin{table}[!ht]
\caption{ }
\begin{tabular}{|c|c|c|c|c|}
\hline $I$ & $|H_{1}H_{2}\rangle$ & $|H_{1}V_{2}\rangle$ & $|V_{1}H_{2}\rangle$ & $|V_{1}V_{2}\rangle$ \\
\hline $F$ & 0.98 & 0.94 & 0.98 & 0.98 \\
\hline\hline $II$ & $|D_{1}D_{2}\rangle$ & $|D_{1}\overline{D_{2}}\rangle$ & $|\overline{D_{1}}D_{2}\rangle$ & $|\overline{D_{1}}\overline{D_{2}}\rangle$ \\
\hline $F$ & 0.97 & 0.96 & 0.95 & 0.99 \\
\hline\hline $III$ & $|R_{1}R_{2}\rangle$ & $|R_{1}L_{2}\rangle$ & $|L_{1}R_{2}\rangle$ & $|L_{1}L_{2}\rangle$ \\
\hline $F$ & 0.95 & 0.95 & 0.96 & 0.97 \\
\hline
\end{tabular}
\end{table}
\par
At the same time the method discussed in this section allows one
to unambiguously distinguish all states forming chosen bases. The
measurement set-up which has been already tested in our
experiments is shown in the Fig. 10.
\begin{figure}[!ht]
\includegraphics[width=0.4\textwidth]{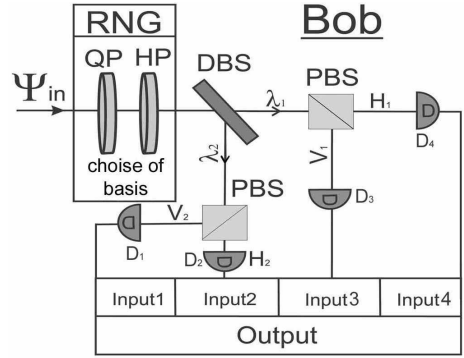}
\caption{Measurement part at Bob's station} \label{setup2}
\end{figure}
It consists of the dichroic mirror, separating the photons with
different wavelenghts, and a pair of polarization beamsplitters,
separating photons with orthogonal polarizations. Four-input
double-coincidence scheme linked  to the outputs of single-photon
detectors registers the biphotons-ququarts. For example for the
first basis, the scheme works as follows, provided that Bob's
guess of basis is correct:
\par if state $|H_{1}H_{2}\rangle$ comes, then detectors D4, D2 will fire,
\par if state $|H_{1}V_{2}\rangle$ comes, then detectors D4, D1 will fire,
\par if state $|V_{1}H_{2}\rangle$ comes, then detectors D3, D2 will fire,
\par if state $|V_{1}V_{2}\rangle$ comes, then detectorsD3, D1 will fire.
 Same holds for any of the remaining correctly guessed bases, since
the quarter- and half waveplates transform the polarization to HV
basis in which the measurement is performed. Registered
coincidence count from a certain pair of detectors contributes to
corresponding diagonal component of the measured density matrix.
So if the basis is guessed correctly, then the registered
coincidence count deterministically identifies the input state. We
illustrate this statement by the table which shows total number of
registered events per 30 sec for the input state
$|R_{1}L_{2}\rangle$ measured in circular basis and calculated
components of experimental (theoretical) density matrix.
\begin{table}[!ht]
\caption{Coincidence rate and density matrix components}
\begin{tabular}{|c|c||c|c||c|c||c|c|}\hline
$D_{4}D_{2}$&$\rho_{11}$&$D_4D_1$&$\rho_{22}$&$D_3D_2$&$\rho_{33}$&$D_3D_1$&$\rho_{44}$\\
\hline
$0$&$0.0(0)$&$220$&$0.973(1)$&$6$&$0.027(0)$&$0$&$0.0(0)$\\
\hline
\end{tabular}
\end{table}
\par
The main obstacle for practical implementation of free-space QKD
protocol based on ququarts is that one needs to perform fast
polarization transformation at the selected wavelengths. There are
different ways how to overcome this problem and we will discuss its
elsewhere. In this section we briefly mention the possible ways.
Since it is not practical to tune the tilting waveplates every time
one wants to encode a quart value, we suggest either to use a
polarization modulator that operates on two wavelengths or to split
the photons with a dichroic mirrors and perform these
transformations on halves of a biphoton independently in a
Mach-Zehnder like configuration. It is important to note that
interferometric accuracy in Mach-Zehnder is not needed, since it is
used only for spatial separation of photons. The practical solution
would be to couple the down converted photons in a single mode fiber
to ensure a perfect spatial mode overlap and then to split them with
wavelength division multiplexer (WDM). Then, the switching between
the states can be done with the polarization modulators that
introduce a $\pi$ or $2\pi$ phase shifts for the selected
wavelength. The choice of basis on Alice's side is done by a zero
order quarter and half waveplates, that can realized within a Pockel
(Liquid Crystal) cell driven by randomly selected voltage. Free
space communication is proposed since it is not practical to
distribute a polarization state within an optical fiber. On Bob's
side, the random choice of basis (RNG) is performed in a same way as
on Alice's side. Then photons are spatially separated with the help
of WDM or dichroic mirror and each of the photon is sent to a
Brown-Twiss scheme with a polarizing beamsplitter that does a
projection of an arrived photon on $H$ or $V$ state as it shown in
Fig. 10. Moreover, registering coincidences allows one to circumvent
the problem of the detection noise that is common for single-photon
based protocols. If the coincidence window is quite small, it is
possible to assure a very low level of accidental coincidences for
the usual dark count rate of single photon detectors.
\par To conclude we have suggested and tested a novel
method of preparation, and measurement of subset of
four-dimensional polarization quantum states. Since for this class
of states the polarization degree is not invariant under SU(2)
transformations it is possible to switch between orthogonal states
using linear transformations like geometrical rotations and phase
shifts. We discussed the family of states that can generate the
whole basis states using only simple polarization elements and
thus might be used in useful applications.
\par While completing this manuscript we learned about closely
related work performed by R. B. R.Adamson and co-authors
\cite{Steinberg}.

\textbf{Acknowledgments}. Stimulating discussions with G.Bj\"ork
and V.P.Karassiov are gratefully acknowledged. This work was
supported in part by Russian Foundation of Basic Research
(projects 05-02-16391a, 06-02-16769, 06-02-16393a) and Leading Russian Scientific Schools (project 4586.2006.2).


\begin{thebibliography}{99}
\bibitem{polardegree} Of course, it is possible to chose
specific orthogonal qutrit states that might be transformed into
each other by geometrical rotations and phase shifts. These states
are: $|HV\rangle$, $|RL\rangle$, and $|D\bar{D}\rangle$. According
to our definition these states posses $P=0$ whereas its have unit
polarization dergee introduced in \cite{gunnar}. In details the
concept of biphoton-qutrits orthogonality and their transformation
properties are discussed in \cite{ortogonality}.
\bibitem{interference} When studying "space-time" interference for polarization-type entangled
states (\ref{eq:prepar1}) the dependence of coinceidence rate on
phase delay $\varphi_{14}$ is analized by setting the polarization
filters at $\pm 45^{\circ}$.
\bibitem{mutual} Bases are mutually unbiased if the inner products
between all possible pairs of vectors belonging to different
bases, have the same magnitude equal to $\frac{1}{\sqrt{D}}$,
where D is the dimension of the Hilbert space.
\bibitem{order} The thickness of the retardation plate that performs the transformation is determined by the number of order $m_{i}$ that is chosen
for the specified wavelength according to
$\frac{m_{1}\lambda{1}}{\delta
n_{1}}=\frac{(2m_{2}+1)\lambda{2}}{2\delta n_{2}}$. Here $\delta
n_{i}$ is the value of the birefringence for the specific
wavelength. We chose order $m_{1}=51$ for $\lambda=702nm$ and
$m_{2}=47$ for $\lambda=650.2nm$.
\end{thebibliography}
\end{document}